\documentclass[aps,prl,twocolumn,groupedaddress,showpacs,showkeys]{revtex4-1}

\usepackage{graphicx}
\usepackage{color}
\usepackage{epstopdf}
\usepackage{subfigure}
\usepackage{amssymb}
\usepackage{amsmath}
\usepackage[latin1]{inputenc}

\begin{document}

\title{ Pure Lattice Gauge Theory in the Expanding Universe }

\author{ Márton Trencséni}
\email{ trencseni@complex.elte.hu }
\affiliation{ Dept. of Physics of Complex Systems, Eötvös University, Pázmány Péter sétány 1/A, 1117 Budapest, Hungary }

\date{\today}

\begin{abstract}
Using Monte Carlo methods, I study the thermodynamic properties of $Z_2$ Abelian lattice gauge theory in flat, homogeneous, isotropic, expanding spacetimes characterized by the scale factor $a(t)$. The presence of the scale factor introduces a spacelike domain wall inside the four dimensional spacetime lattice where the phase transition associated with deconfinement occurs. The resulting theory appears identical to introducing a time-dependent effective coupling $\beta_{\textrm{eff}}$, and could serve as a simple model for the electroweak phase transition of the early Universe.
\end{abstract}


\maketitle

Lattice gauge theory, the study of gauge theories on discretized spacetimes is an important tool for the study of quantum theories such as QCD and the Standard Model. On the lattice path integrals associated with the theory become finite dimensional, making it is possible to approximate them on the computer using Monte Carlo methods. Lattice gauge theories were introduced by Wilson's 1974 definition of a discretized action with the appropriate continuum limit \cite{Wilson}. Creutz' series of papers in the following years investigated the thermodynamic properties of $Z_N$, $U(1)$ and $SU(N)$ pure lattice gauge theories using computer simulations, and found that the theories contain phase transitions associated with deconfinement \cite{CreutzAbelian, CreutzSU2, CreutzSU5, CreutzSU6}. This demonstrated that gauge theories are a viable framework for the experimentally observed phenomenon of quark confinement.

These, and subsequent investigations that followed were carried out in Euclidean spacetime. In the context of this letter, it is instructive to think of Euclidean spacetime as a Wick rotated Minkowski spacetime, where Wick rotation is defined by the replacement $t  \rightarrow it$. Wick rotation is necessary to get rid of the imaginary exponent in the original path integral, to arrive at a theory which can be simulated effectively on a computer. But the results of the computational experiments describe the physics of gauge theories in Minkowski spacetime.

Standard cosmological theory, confirmed recently the Wilkinson Microwave Anisotropy Probe \cite{WMAP7} and the Sloan Digital Sky Survey \cite{SDSS}, tells us that the Universe is expanding, not static. On cosmological scales the Universe is described by the spatially flat, homogeneous and isotropic Friedmann-Lemaitre-Robertson-Walker (FLRW) metric, which contains a scale factor $a(t)$ to describe the expansion of space with cosmological time.

In this letter, I define pure lattice gauge theories on Euclidean FLRW metrics and present results of computational simulations using the simplest $Z_2$ Abelian model. These demonstrate that the presence of the scale factor introduces a spacelike domain wall inside the four dimensional spacetime lattice where the phase transition associated with deconfinement occurs. This indicates that even simple lattice gauge theory models extended with the scale factor $a(t)$ can qualitatively reproduce the electroweak phase transition of the early Universe.

\begin{figure}[floatfix]
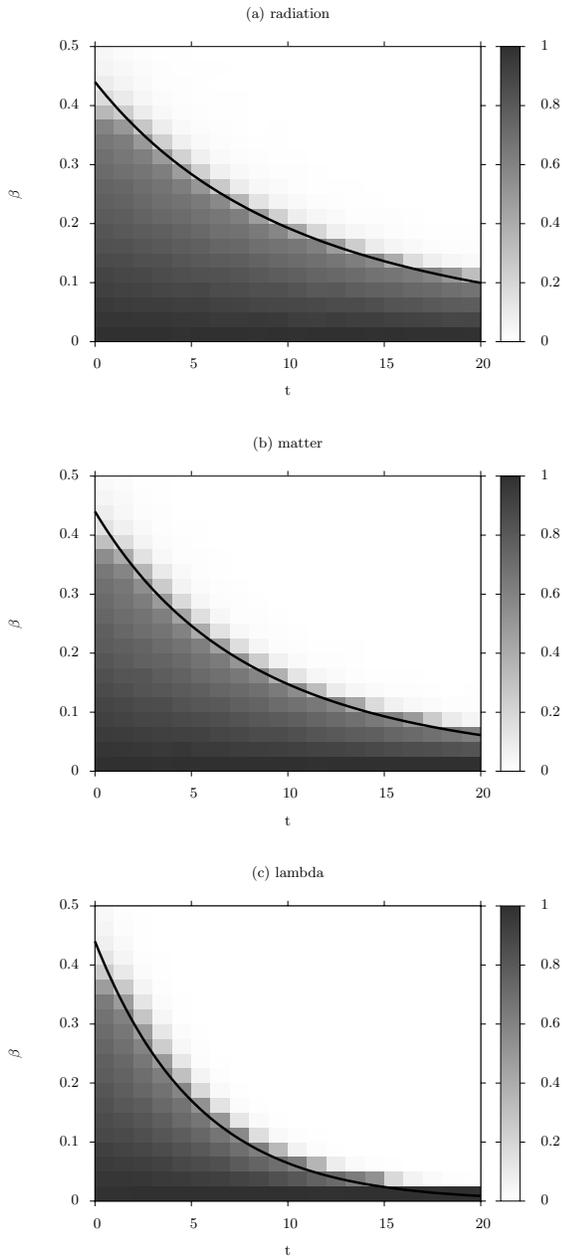

 \subfigure{
  \resizebox{8cm}{!} { \input{radiation.tex} }
 }
 \subfigure{
  \resizebox{8cm}{!} { \input{matter.tex} }
  }
 \subfigure{
  \resizebox{8cm}{!} { \input{lambda.tex} }
}
\caption{\label{CPT}Results of Monte Carlo runs showing the order parameter $E$ for an effective $20^4$ lattice with different cosmological scale factors. The black line corresponds to $\beta_{\textrm{eff}}(t) = \beta_c$. The plots demonstrate that a cosmological phase transition occurs when this effective critical inverse temperature is reached.}
\end{figure}

I formulate the theory on a four-dimensional hypercubic lattice. Associated with each link joining a pair of nearest-neighbor sites $i$ and $j$ is an element $U_{ij}$ of the group $Z_2 = \{+1, -1\}$, oriented such that $U_{ij} = U^{-1}_{ji}$. The discretized action describing these spins is $ S = \sum_{ \square } S_{ \square } $, where the sum is over all elementary spacetime plaquettes, and $S_{\square}$ is the per plaquette action defined as
$$ S_{\square} = \sqrt{|g|} \; g_{\mu \mu} g_{\nu \nu} \left(1 - U_{ij} U_{jk}  U_{kl} U_{li} \right) $$
for a plaquette in oriented in the $\mu \nu$ plane, where $g_{\alpha\beta}$ is the Euclidean FLRW metric $ \left( g_{00} = 1, \; g_{ii} = a^2(t) \right)$ for the chosen cosmology, and $g = \det(g_{\alpha\beta}) = a^6(t) $. The extra $g$ terms in this expression are there to reproduce the Einstein-Hilbert action in the continuum limit, which for flat spacetime with matter is $ S_{EH} = \int \mathcal L \; \sqrt{|g|} \; d^4x $.

To see this, note that the Lagrangian density for the Abelian gauge field $ F_{\mu \nu} = \partial_{\mu} A_{\nu} - \partial_{\nu} A_{\mu} $ is
$$ \mathcal L = - \frac{1}{2} F_{\mu \nu} F^{\mu \nu} = - \frac{1}{2} g^{\alpha \gamma} g^{\beta \delta} F_{\alpha \beta} F_{\gamma \delta} $$
where the group elements are connected to the vector potential $A_{\mu}$ by $ U_{ij} = e^{i g_0 A_{\mu} d } $ where $g_0$ is the coupling constant of the theory, $\mu$ is the direction of the bond specified by $ij$, and $d$ is the lattice spacing.

We insert the discretized action into a path integral to define a partition function at inverse temperature $\beta = 2 / g_0^2$,
$$ Z  = \ \sum_{\{ U_{ij} \}} e^{- \beta S} $$
where the sum runs over all possible values of the link variables $U_{ij}$. The order parameter used to search for phase transitions is the average action per plaquette
$$ E = \langle 1 - U_{ij} U_{jk}  U_{kl} U_{li} \rangle $$
Unlike in traditional pure lattice gauge simulations where the order parameter is computed for the entire four-dimensional spacetime lattice, I compute and plot $E$ for each time slice. Traditional lattice gauge theory is recovered if $a(t) = 1$ is used in the equations above, in which case $E$ does not have time-dependence.

For brevity the discussion above is specific to the $Z_2$ gauge group, as this is the simplest model to work with computationally, but produces the novelties of introducing a scale factor. However, the framework for including a scale factor in a pure lattice gauge theory discussed above works in general for $Z_N, U(1)$ and non-Abelian gauge groups $SU(N)$.

The partition function $Z$ still contains a large number of terms, $2^{640000}$ for a $20^4$ $Z_2$ lattice, and direct evaluation on the computer is not feasible. The Monte Carlo method is used where a sequence of configurations is generated which estimates the full sum, at thermal equilibrium at inverse temperature $\beta$, using the heat bath method \cite{CreutzAbelian}. This procedure selects a new value $U_{ij}$ for each spin variable in a stochastic manner with probability distribution proportional to the Boltzmann factor $ e^{- \beta S} $, conforming to the condition of detailed balance, assuring that the procedure transforms an ensemble in equilibrium into itself.

After setting up the lattice gauge framework, I discuss the cosmological background used in the simulations. In accordance with astrophysical measurements, on cosmological scales our Universe is described by the spatially flat, homogeneous and isotropic FLRW metric $ ds^2 = a^2(t) d\Sigma^2 - dt^2 $, where $ d\Sigma^2 = d^2x + d^2y + d^2z $.

With the FLRW metric the Einstein equations, describing the interaction of spacetime and matter, reduce to the Friedmann equations, a set of second order differential equations which can be solved for $a(t)$ if an equation of state $p = p(\rho)$ is specified, where $p$ is pressure and $\rho$ is energy density. In cosmology the perfect fluid equation of state $ p = w \rho $ is used, characterized by the dimensionless number $w$. Different values of $w$ correspond to different source types.

In this Letter I work with three source types commonly considered in cosmology: matter, radiation and lambda.  Table  ~\ref{TableScaleFactors} gives $w$ and the form of the scale factor $a(t)$ arising from the Friedmann equations for these source types.

\begin{table}[htdp]
\caption{\label{TableScaleFactors}Scale factor for standard source types.}
\begin{ruledtabular}
\begin{tabular}{l c c}
        source      & $w$   & $a(t)$ \\
        \hline \\ [-1.5ex]
        matter      & $0$   & $t^{\frac{2}{3}}$ \\
        radiation   & $1/3$ & $t^{\frac{1}{2}}$ \\
        lambda      & $-1$  & $exp(t)$  \\
\end{tabular}
\end{ruledtabular}
\end{table}

The simulations are carried out on a $40 \times 20^3$ hypercubic lattice using periodic boundary conditions. I let the spacetime expand until $T_h = 20$ and then symmetrically contract it back by setting $a(T_h + t) = a(T_h - t)$ so that it can be wrapped onto itself at $T = 40$. This gives an effective lattice size of $20^4$, as the thermodynamic quantities of the two halves are mirror images of each other. Only the first, expanding half is shown in the following figures.

I run simulations using different $\beta$s, starting at $\beta = 0$ to $\beta = 0.5$ at $ \Delta \beta = 0.025 $ intervals. Before each $\beta$-run, the lattice is initialized to $U_{ij}$ = 1. I experimented with different initialization strategies, such as completely disordered states, but these did not affect the overall results. For each $\beta$-run, the entire lattice is updated 100 times using the Monte Carlo heat bath algorithm, a value which was found to give good convergence results. The size of the lattice, $\beta$-resolution and number of Monte Carlo iterations used is only limited by computational resources. Comparing my results with past results \cite{CreutzAbelian} shows that these values are sufficient for good convergence and understanding the thermodynamic properties of the system. With these values, a simulation run takes a few minutes on a standard desktop computer.

\begin{figure}[floatfix]
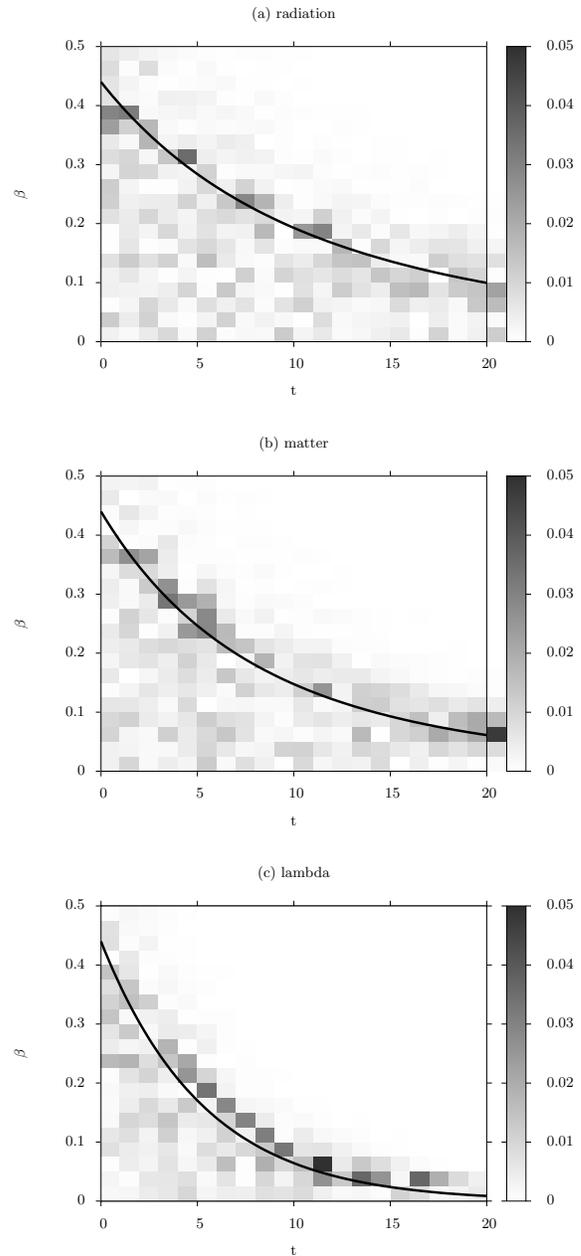

 \subfigure{
  \resizebox{8cm}{!} { \input{diff_radiation.tex} }
 }
 \subfigure{
  \resizebox{8cm}{!} { \input{diff_matter.tex} }
 }
 \subfigure{
  \resizebox{8cm}{!} { \input{diff_lambda.tex} }
 }
\caption{\label{DiffMap}Difference in the order parameter $E$ for an effective $20^4$ lattice between Monte Carlo simulations where a cosmological scale factors was used versus a time-dependent $\beta_{\textrm{eff}}(t) = \beta \left[ a^5(t) + a^7(t) \right] / 2$. The black line corresponds to $\beta_{\textrm{eff}}(t) = \beta_c$. The plots demonstrate that the statistics of the system is well approximated by the time-dependent effective $\beta$. Since $\beta = 2 / g_0$, this corresponds to a time-dependent coupling constant in the gauge theory.}
\end{figure}

Figure ~\ref{CPT} shows the results of simulation runs for the three source types considered. A pixel on each of these 2D maps is the average action $E$ of the 4D lattice for a simulation run at a given $\beta$ (row), at each time step $t$ (column) after the 100 Monte Carlo iterations have completed. Since I am working in 4D spacetime, the entire row is extracted at once at the end of each $\beta$-run. For a fixed $\beta$ below $\beta_c = 0.44$, the Universe starts off in a completely disordered state, at some $\beta$ dependent time reaches a phase transition and then becomes completely ordered. Examining the three cases (a) - (c), the transition occurs sooner for cosmologies where the scale factor $a(t)$ grows faster, lambda being the fastest and radiation the slowest. Above $\beta_c$, the Universe is  ordered at all times and no phase transition occurs inside the spacetime lattice. The critical value $\beta_c = \frac{1}{2} \ln( 1 + \sqrt{2} ) \approx 0.44 $ is known from the self-duality of the $Z_2$ theory \cite{BetaCrit}.

The black line on all three plots corresponds to $\beta_{\textrm{eff}}(t) = \beta \left[ a^5(t) + a^7(t) \right] / 2 = \beta_c$, and is displayed to show that although different scale factors are employed in the three simulations, there is a common theme. The presence of the scale factor introduces the effective inverse temperature $\beta_{\textrm{eff}}(t)$, whose form is explained by inspecting the expression for the per plaquette action $S_{\square}$. The metric factors in $S_{\square}$ give either $a^5(t)$ or $a^7(t)$, the former for timelike plaquettes where either $\mu$ or $\nu$ is 0, the latter for spacelike plaquettes where neither $\mu$ or $\nu$ is 0, and there are an equal number of timelike and spacelike plaquettes in a hypercubic lattice.

To verify that the cosmological model with a scale factor is statistically equivalent to one with a time-dependent effective $\beta$, I have performed simulations on the same lattice, removing the scale factor but letting $\beta = \beta_{\textrm{eff}}(t)$ vary with time according to the functional forms given in Table ~\ref{TableScaleFactors}. In this approximation both timelike and spacelike per-plaquette actions are computed at the same $\beta_{\textrm{eff}}(t)$ inverse temperature at each time slice. I have subtracted these maps from the cosmological maps to examine the error of this estimation. The results are shown on Fig. ~\ref{DiffMap} and demonstrate good agreement, with the largest deviation occurring at the phase transition around the $\beta_{\textrm{eff}}(t) = \beta_c$ line, when the system in undergoing the largest fluctuations.

\begin{figure}[floatfix]
\resizebox{8cm}{!} { \input{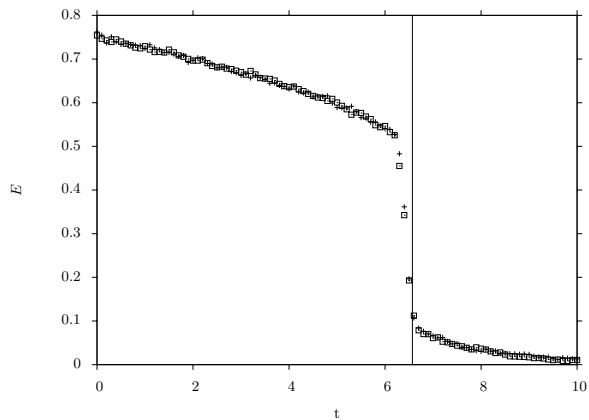} }
\caption{\label{FixedBeta}$E$ for high resolution Monte Carlo runs for cosmology (crosses) and time-dependent effective $\beta$ (squares), for fixed $\beta = 0.25$. In both cases the scale factor $a(t)$ for radiation is used. The vertical line corresponds to $t=6.58$ where $\beta_{\textrm{eff}}(t)=\beta_c$.}
\end{figure}

To further examine the $\beta_{\textrm{eff}}(t)$ approximation, I have run high resolution simulations at fixed $\beta = 0.25$ for the radiation case, with both the cosmological scale factor and the effective inverse temperate on a lattice with effective size $200 \times 20^3$ using $100$ Monte Carlo iterations. Results are shown in Fig. ~\ref{FixedBeta} and show excellent agreement between the two models. At $t = 0$ the order parameter starts out at $E \approx 0.75$, the value for $E$ when performing a traditional $Z_2$ 4D simulation at $\beta = 0.25$. The vertical line is the critical $\beta_c$ value at which the phase transition is predicted to occur. As first shown by Creutz \cite{CreutzAbelian}, Fig. ~\ref{FixedBeta} demonstrates that the $Z_2$ model is first-order discontinuous with its sharp drop at $\beta_c$.

The traditional $Z_2$ model shows hysteresis in $E$ when a thermal cycle is simulated on the system, which corresponds to first increasing $\beta$ from $0$ to above $\beta_c$, and then back to $0$. For the system studied in this letter, the effective $\beta$ is a monotonic function of $a(t)$ in the relevant region $a(t) > 1$. The simulation is set up in such a way that $a(t)$ increases in the first half of the lattice, for $t < T_h$, and then decreases symmetrically for $t > T_h$, so periodic boundary conditions can be used. However, comparing the two sides of the lattice I found no hysteresis effect when flipping over $E$ about $T_h$. Hysteresis occurs in the original model because as $\beta$ is increased (or decreased) for each iteration, the lattice is in the state of the previous iteration, and this dependence results in different curves for the two $\beta$ directions. A quick modification to the cosmological simulation verifies that hysteresis can be reintroduced by setting the $U_{ij}$ variables at time slice $t_{i+1}$ equal to the corresponding variables at $t_i$ when performing the simulation.

I plan to study this modified system, along with the more general $Z_N$, $U(1)$ and $SU(N)$ gauge groups in the future.

I would like to thank Michael Creutz and István Csabai for helpful comments and reviews of early drafts of this Letter.

\end{document}